\documentclass[10pt]{iopart}
\usepackage{iopams}
\usepackage[dvips]{graphicx}

\begin{document}

\newcommand{\degre}{\(\mathsurround=0pt{}^\circ\)}

\title{Thermally Activated Dynamics of the Capillary Condensation}

\author{Frédéric  Restagno \footnote[3]{To
whom correspondence should be addressed. E-mail address:  frestagn@ens-lyon.fr} \dag, Lydéric Bocquet \dag\ Thierry Biben \dag and \'Elisabeth Charlaix \ddag}
\address{\dag\ Laboratoire de Physique (UMR CNRS 5672, \'Ecole Normale Supérieure de Lyon, 46 allée d'Italie, 69364 Lyon Cedex 07, France, }
\address{\ddag\ Département de Physique des Matériaux (UMR CNRS 5586),
    Université Lyon~I, 69622 Villeurbanne cedex, France}

\begin{abstract}
This paper is devoted to the thermally activated  dynamics of the
capillary condensation.  We present a simple model which enables
us to identify the critical nucleus involved in the transition mechanism. This simple model 
is then applied to calculate the nucleation barrier from which 
we can obtain informations on the nucleation time. We present a simple estimation of the nucleation barrier in slab geometry both in the two dimensional case and in the three dimensional case. We extend the model in the case of rough surfaces which is closer to the experimental case and allows comparison with experimental datas.
\end{abstract}

\submitted

\maketitle{}

\section{Introduction}
When two surfaces are brought together in a condensable vapor near saturation, a first order phase transition from gas to liquid occurs at small gap width provided that the solid wets the solid substrate {\i.e.} has a contact angle smaller than 90\degre.
Macroscopic considerations 
predict that the condensation occurs for distance between the solid
surfaces $H$ less than a critical distance $H_c$ verifying 
\begin{equation}
\Delta\rho~\Delta\mu \simeq {{2(\gamma_{SV}-\gamma_{SL})} / H_c}
\label{Kelvin}
\end{equation}
where $\Delta\rho=\rho_l-\rho_g$ is difference between the bulk densities
of the liquid and the gas and $\Delta\mu=\mu_{sat}-\mu$ is the (positive) 
undersaturation
in chemical potential, with $\mu_{sat}$ the chemical potential at bulk
coexistence \cite{Israelachvili85}.  If the vapor is assumed to be a perfect gas, then $\Delta \mu \approx k_BT\ln(P_{sat}/P_{vap})=k_BT\ln(1/RH)$, where $RH$ is the the so-called relative humidity. At standard ambient conditions for water ($\gamma_{LV}=72 \ \textrm{mJ.m}^{-2}$, $\rho_L\approx 3.10^{28}\ \textrm{m}^{-3}$, $H=40\%$), we obtain $H_c\approx 2 \ \textrm{nm}$. Capillary condensation is usually invoked to interpret adsorption isotherms of gases in mesoporous media \cite{Perez98}. This transition is now well understood and documented, both
from the experimental \cite{Crassous94,Israelachvili79,Christenson84} and theoretical point of view 
\cite{Evans86,Evans89}. 

On the other hand, the problem of the {\em dynamics} of the transition has receive very little attention.  Experimentally, only indirect 
informations on the dynamics are available in the litterature. Experimental studies of
capillary condensation using the Surface Force Apparatus (SFA) technique
systematically show an important hysteresis in the interaction force between 
two substrates when the separation of the surfaces is first decreased and then increased. This large hysteresis points out the strong metastability of the gas phase when $H<H_c$, which persists over macroscopic times. Recent experiments with SFA have studied the growth of the liquid meniscus after the nucleation \cite{Kohonen99} but no attention has been given to the nucleation time which is much larger than this growing time.
Experiments measuring the cohesion inside divided materials may provide indirect information 
on the dynamics of the transition too \cite{Bocquet98,Restagno99,Crassous99}. Theoretically, lattice-gas simulations showed that the
(topologically equivalent) drying transition occurs via the creation of
``tubes'' connecting the two wetting films \cite{Lum97}.

Beyond these results, a theory proposing a mechanism for the 
dynamics of the capillary condensation is still needed.

In the following, we show how to construct the critical nucleus 
for capillary condensation. First, a simplified model in the slab geometry
keeping only the main ingredients for capillary condensation will be
considered. The latter has both advantages to allow tractable calculations and to capture 
the essential features of the involved physics. In a second part, we show how the natural roughness of the surfaces on a nanometric scale can be taken into account on the dynamics of the transition. Applications to the adsorption kinetics in a granular medium shall be discussed.

\section{The slab geometry}

In a first step, we restrict our attention to a system confined between two perfectly smooth and flat solid surfaces, and in contact with a reservoir of temperature $T$ and chemical potential $\mu$.

 Let us consider the situation in which planar liquid films
 of varying thickness $e$ ($e<H/2$) develop on both solid surfaces. 
 Following Evans {\it et al.} \cite{Evans85,Evans86}, the grand potential of the system may be written
 \begin{equation}
 \Omega=-p_V V_V-p_L V_L+2\gamma_{SL} A+2\gamma_{LV} A
 \label{omega}
 \end{equation}
 where $V_V$ (resp. $V_L$) is the volume of the gas (resp. liquid) phase and
 $A$ is the surface area.
 Using $V_L=2Ae$, $V_V=A(H-2e)$ and $p_V-p_L\simeq \Delta\rho\Delta\mu$, one gets
 \begin{equation}
 \Delta\omega(e)\equiv{1\over A}(\Omega-\Omega(e=0))=\Delta\rho\Delta\mu~ 2e
 \label{domega}
 \end{equation}
 Note that in the complete wetting situation $\Omega(e=0)$ can be identified
 with $\Omega_V$, the grand-potential of the system filled with the gas
 phase only. The situation $e=H/2$ corresponds to the opposite case where
 the two liquid films merge to fill the pore. The grand potential thus exhibits
 a discontinuity in $e=H/2$ corresponding to the disappearance of the two
 liquid-vapor interfaces, and its value is reduced by $2\gamma_{LV}A$.
 When $e=H/2$, expression (\ref{domega}) must then be replaced by 
 $\Delta\omega(e=H/2)=-\Delta\rho\Delta\mu (H_c-H)$, where $H_c$ is the critical
 distance defined in eq. (\ref{Kelvin}).  One may note
 that the minimum of the grand potential corresponds to a complete filling
 of the pore by the liquid phase when $H<H_c$, as expected. If we now
 allow deformation of the interfaces, {\it i.e.} the thickness $e$ is now a
 function of the lateral coordinates, the corresponding cost has to be added
 to the grand potential. We assume also a mirror symetry of the interfaces, so that one finds in this case:
 \begin{equation}
 \Delta\Omega_{tot}= \gamma_{LV} \Delta A_{LV}+\int dS \Delta \omega(e)
 \label{eq1}
 \end{equation}
 with $\Delta\Omega_{tot}=\Omega(\{e\})-\Omega_V$ and $\Delta A_{LV}=A_{LV}-A$
 is the excess $L-V$ area. The integration in the last term runs over
 the solid surface.

\subsection{The 2D case}
\begin{figure}[htbp]
\begin{center}
\includegraphics[width=100mm]{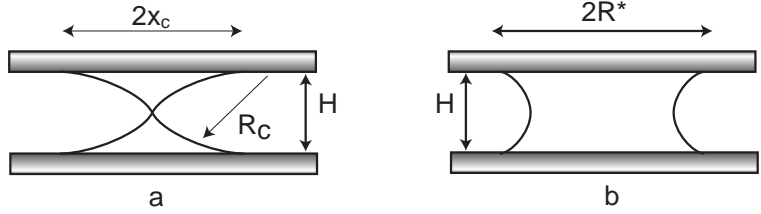}
\caption{a : Picture of the critical nucleus for capillary condensation in two dimensions and perfect 
wetting case ($\theta=0$). The radius of curvature of the
meniscus is equal to $R_c=H_c/2$, which is only approximatively verified
within the small slope assumption. See text for details. b: Picture of the critical nucleus in three dimensions and perfect wetting case. See text for details.}
\label{fig1}
\end{center}
\end{figure}

Let us consider first the {\bf 2D case}. Within the small slope 
 assumption, $\Delta A_{LV} \simeq \int dx~{\gamma_{LV}}\vert\nabla e\vert^2$,
 extremalization of the grand potential 
 leads to the following Euler-Lagrange equation for $e(x)$, where $x$ denotes
 the lateral coordinate:
 \begin{equation}
 2\gamma_{LV} {d^2e\over{dx^2}} - {d\Delta \omega(e)\over de} =0
 \label{eq2}
 \end{equation}
 We look for solutions satisfying $e=0$ and
 $de/dx=0$ at infinity. We can choose $e(x=0)=H/2$ to
 fix the origin. The complete solution, depicted in fig
 \ref{fig1}.a, can be obtained in the form of parabolic branches with a   
 spatial extension
 $x_c=\sqrt{H R_c}$ where $R_c=H_c/2$.  
 Let us note that the cusp in the solution in $x=0$ stems from 
 the discontinuity of $\Delta\omega$ in $e=H/2$ resulting from the assumption
 of an infinitesimely narrow liquid-vapor interface. 
 Condensation thus occurs through the
 exitation of short wavelength fluctuations, in agreement with the simulations
 results for the drying transition \cite{Lum97}.
 The corresponding energy of the nucleus (per unit length in the perpendicular direction) 
 can be calculated by integration of eq. (\ref{eq1}):
 \begin{equation}
 \Delta \Omega^{\dag} = {4\over 3} (\Delta\mu \Delta\rho \gamma_{LV})^{1/2}
 H^{3/2}
 \label{eq6}
 \end{equation}

 It is easy to check that $\Delta \Omega^{\dag} $ corresponds to a saddle-point of
 the grand-potential. It is greater than both free energies of the gas and
 liquid phases. Moreover $\Delta \Omega^{\dag} $ is smaller than the 
 free energy of any other configuration maximizing the grand potential 
 since it is the only solution of finite extension. 
 We just point out that the parabolic solution obtained above
 is the small slope approximation to the circle with radius of curvature $R_c$.

We mention that the prediction for $\Delta \Omega^{\dag}$ in eq. \ref{eq6} is in agreement with numerical simulations results, using a Landau-Ginzburg model for the grand potential of the system together with a non conserved Langevin dynamics. Full details of these simulations are given elsewhere \cite{Bocquet99}.

These results can be generalized to the partial wetting case. The only difference is that the contact angle $\theta$ on the surfaces is now non vanishing and obeys Young's law, $\gamma_{SV}-\gamma_{SL}=\gamma_{LV}\cos\theta$. To leading order on $H/H_c$, one gets \cite{Bocquet99}~:
\begin{equation}
\Delta \Omega^{\dag} \simeq 2 \gamma_{LV}\sin \theta H
\end{equation}

\subsection{The 3D case}

The previous approach can now be directly generalized to the {\bf 3D case}.
 The physical understanding of the results can be however greatly simplified
 if one realizes that 
 {\it maximization} of the grand potential, eq. (\ref{eq1}), leads to two   
 {\it mechanical equilibrium} conditions : the usual Laplace
 equation, relating the local curvature $\kappa$ to the pressure drop
 ${\gamma_{LV} \kappa}=\Delta p\simeq \Delta \mu\Delta \rho$; and the Young's 
 law which fixes the contact angle of the meniscus on the solid substrate
 according to $\gamma_{LV} \cos\theta=\gamma_{SV}-\gamma_{SL}$ ({\it i.e.} 
 $\theta=0$ in the perfect wetting case).
 These non-linear equations cannot be
 solved analytically in 3D, but one can easily understand that the corresponding
 critical nucleus takes the form of a liquid bridge of finite lateral extension $R^*$,
 connecting the two solid surfaces (see fig \ref{fig1}b). This finite extension
 results physically
 from the balance between a ``surface'' contribution $\Delta\Omega_1 \approx 
 (\Delta\rho\Delta\mu H-2(\gamma_{SV}-\gamma_{SL})) \pi R^2$ which 
 drives capillary condensation, and a linear contribution 
 $\Delta\Omega_2  \approx 2\pi\gamma_{LV} R H$ 
 specific to the 3D case which tends to close the bridge. Maximization of the 
 free energy gives a finite extension $R^*$, yielding for the free energy  
 barrier :
 \begin{equation}
\Delta \Omega^\dag \approx {\pi \gamma _{LV}^2\over {2(\gamma_{SV}-\gamma_{SL})}} 
{H^2 H_c \over {H_c-H}}	
 \label{NRJbarrier3D}
 \end{equation}
 Full details in the 3D case shall be given in a forthcoming paper
\cite{Bocquet99b}.

\section{The rough case}

\begin{figure}[htbp]
\begin{center}
\includegraphics{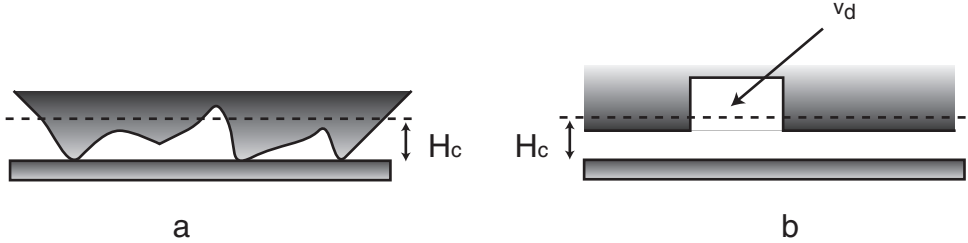}
\caption{a : Typical representation of two rough surfaces. Note that we consider the roughness at the nanometric scale. \\  b  : Schematic representation of an asperity. $v_d$ is the excess volume of the defect, $a_d$ the area of the defect.}
\label{fig2}
\end{center}
\end{figure}

\begin{figure}[htbp]
\begin{center}
\includegraphics{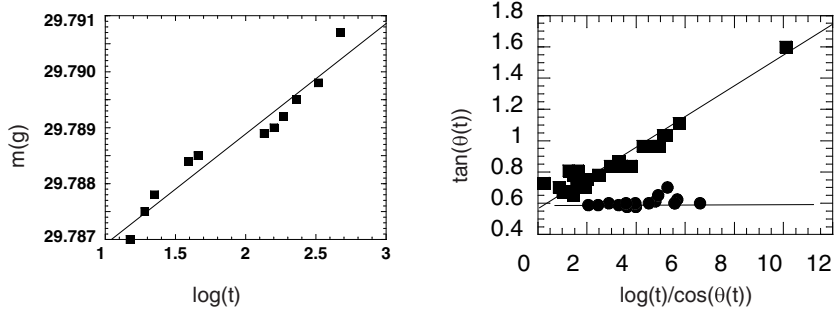}
\caption{a : Evolution of the mass $m$ of a pile of glass beads of radius smaller than 50 $\mu$m as a funtion of the logarithm of the resting time $t$ in hours. Note that the time is comprised between a few minutes and  2 weeks. The temperature is fixed at $31\pm 0.1$ \degre C. The relative humidity is fixed at $68 \% $ by the salt method described in reference \cite{Restagno99}. The straight line is the best linear fit of the datas.  b : Evolution of the tangente of the maximum stability angle $\theta_m$ of a an assembly of glass beads as a function of the logarithm of the "resting" time $t$ in seconds (divided by the cosine of this angle which comes from geometrical arguments). This angle is measured in a cylinder. The full experimental setup is described in \cite{Restagno99}. (\fullcircle) : $RH=3\%$, (\fullsquare) : $RH=43 \% $. The straight lines are the best linear fits of the datas.}
\label{fig3}
\end{center}
\end{figure}

Although a lot can be learned from the perfectly flat slab geometry, the latter is certainly too idealized to account for the kinetics of adsorption in "real" experimental systems. In particular, very slow logarithmic depedence have been measured on various static properties of granular media in the presence of humidity (see fig \ref{fig3} and text hereafter) \cite{Bocquet98, Restagno99}.
  As we shall show hereafter, these logarithmic time dependence may be understood by taking into account the influence of roughness on the dynamics of capillary condensation. Let us consider a simple model consisting of two surfaces facing each other 
and rough  at the nanometric scale, as depicted on fig. 
(\ref{fig2}.a). As emphasized in the introduction capillary condensation 
typically occurs in pores of nanometric size. We thus have to  consider the roughness of 
the surfaces at the {\it nanometer level}.
Without loss of generality,
one may consider that one of the walls is perfectly flat. 
When roughness is present, there is a broad range of gaps between the surfaces. 
In particular, there are regions where the two surfaces are in close contact. In such regions, condensation should take place on a very short time-scale. Thus at ``early times'', one has to consider a set of wetted islands, which we shall 
consider as independent. Once these islands have formed, they should grow up to 
a point where the distance between the surfaces is equal to $H_c$, so that a 
meniscus of radius $R_c=H_c/2\cos\theta$ forms at the liquid-vapor interface, 
allowing for mechanical equilibrium. 

In doing so however, the wetted area has to overcome unfavorable regions where the 
distance between the two surfaces is larger then $H_c$. Let us consider a 
specific jump over such a ``defect'', as
idealized in fig. (\ref{fig2}.b). We denote by $e_d$ the ``averaged'' gap 
inside the defect ($e_d>H_c$), and by $a_d$ its area.
The free energy cost for the liquid bridge to overcome this defect is 
approximatively given by
\begin{eqnarray}
&\Delta \Omega^{\dag} &\simeq a_d \left(\Delta\mu\Delta\rho~e_d-2\Delta\gamma\right)\nonumber\\
& & \equiv v_d \Delta\mu\Delta\rho 
\label{CCR1}
\end{eqnarray}
where $v_d$ is the excess volume of the defect, $v_d=a_d~(e_d-H_c)$.
We can thus estimate the time to overcome the defect as
\begin{equation}
\tau=\tau_0 exp \left\{ {\Delta \Omega^{\dag} \over {k_BT}}\right\}
\label{CCR2}
\end{equation}

One may expect the defects  to exhibit a broad distribution of excess volume  $v_d$, so that the activation times $\tau$ are accordingly widely distributed.  After a time $t$, only the defects with an activation time $\tau$ smaller than $t$ have been overcomed. Using eq. \ref{CCR1} and \ref{CCR2} these have an excess volume  $v_d$ which verifies $v_d<v_{dmax}(t)=k_BT(\Delta\mu\Delta\rho)^{-1}\ln(t/\tau_0)$. At a time $t$, the number of overcomed defects is then typically $N(t)=v_{dmax}(t)/v_0$ where $v_0$ is the typical width of the distribution of excess volume of the defects. Now, once a liquid bridge has bypassed a defect, it fills locally the volume surrounding the nucleating site and the wetted area increases by some typical (roughness dependent) amount $\delta A_0$. The time dependent wetted area can thus be written~:
\begin{equation}
A_w(t) \simeq N(t)\delta A_0 =\frac{\delta A_0}{\Delta\mu/(k_BT)\Delta\rho v_0}\ln\left(\frac{t}{\tau_0}\right)
\label{equ10}
\end{equation}

Similar expressions with logarithmic dependence on time, can be found on other quantities, like the time dependent adsorbed amount, or the adhesion force between rough surfaces.

These logarithmic dependence have been observed in two kinds of experiments. In the first one (see fig. \ref{fig3}.a), we have measured the evolution of the mass  of a sample of glass beads with a diameter smaller than 50 micrometers at fixed humidity ($RH = 68\%$) as a function of the resting time $t$ \cite{Restagno2000}. The glass beads where first dried at high temperature and put at the fixed humidity controlled by the saturated-salt method described in \cite{Restagno99}. The evolution of the mass fits well with a logarithmic behavior, as described on eq. \ref{equ10}. On the other hand, the cohesion force resulting from condensation of liquid bridges has been probed in a granular medium by measuring the maximum angle of stability as a function of resting time. As shown on figure \ref{fig3}.b, the latter exhibits a slow logarithmic dependence in agreement with eq. \ref{equ10} \cite{Bocquet98,Restagno99}.

The authors would like to thank J. Crassous, J.C. Geminard and H. Gayvallet
for many interesting discussions. This work has been partly supported by
the PSMN at ENS-Lyon, the MENRT under contract 98B0316 and the franco-british 
program ALLIANCE (contract 99041)


\vspace{0.5 cm}
\noindent {\bf References}
\vspace{0.5 cm}
\begin{thebibliography}{10}

\bibitem{Israelachvili85}
Israelachvili J 1985 {\em Intermolecular and Surfaces Forces\/} (London: Academic Press)

\bibitem{Perez98}
P{\'e}rez L, Sokolowski S and Pizio O 1998 {\em J. Chem. Phys.\/} {\bf 109} 1147--1151

\bibitem{Crassous94}
Crassous J, Charlaix E and Loubet J 1994 {\em Europhys. Lett.\/} {\bf 28} 37--42

\bibitem{Israelachvili79}
Israelachvili J~N 1979 {\em Nature\/} {\bf 277} 548--549

\bibitem{Christenson84}
Christenson H~K 1984 {\em J. Colloid Interface Sci.\/} {\bf 104} 234--249

\bibitem{Evans86}
Evans R, Marconi U~M~B and Tarazona P 1986 {\em J. Chem. Phys.\/} {\bf 84} 2376--2399

\bibitem{Evans89}
Evans R 1989 {\em Liquids and Interfaces\/}, edited by J~Charvolin, J~Joanny and J~Zinn-Justin (Elsevier Science Publishers)

\bibitem{Kohonen99}
Kohonen M, Maeda N and Christenson H 1999 {\em Phys. Rev. Lett.\/} {\bf 82} 4667--4670

\bibitem{Bocquet98}
Bocquet L, Charlaix E, Ciliberto S and Crassous J 1998 {\em Nature\/} {\bf 396} 735

\bibitem{Restagno99}
Restagno F, Gayvallet H, Bocquet L and Charlaix E 1999 {\em Dynamics in Small Confining Systems {IV}\/}, edited by J~Drake, G~Grest, J~Klafter and R~Kopelman MRS (Boston: MRS) vol. 543

\bibitem{Crassous99}
Crassous J, Bocquet L, Ciliberto S and Laroche C {\em Europhys. Lett.\/} {\bf (in press)}

\bibitem{Lum97}
Lum K and Luzar A 1997 {\em Phys. Rev. E\/} {\bf 56} R6283--R6286

\bibitem{Evans85}
Evans R, Marini U and Marconi B 1985 {\em Chem. Phys. Lett.\/} {\bf 114} 415

\bibitem{Bocquet99}
Bocquet L, Restagno F and Biben T (see cond-mat/9901180 ) {\em Submitted to  Phys. Rev. Lett.\/}

\bibitem{Bocquet99b}
Bocquet L, Restagno F, Charlaix E and Biben T {\em in preparation\/}

\bibitem{Restagno2000}
Restagno F, Gayvallet H, Bocquet L and Charlaix E {\em In preparation\/}

\end{thebibliography}
\end{document}